\documentclass[prl,superscriptaddress,twocolumn]{revtex4}
\usepackage{enumerate}
\usepackage{amsfonts,amssymb,amsmath}
\usepackage[]{graphics,graphicx,epsfig}
\usepackage{amsthm}

\usepackage{marvosym}

\usepackage{svg}
\usepackage{fdsymbol}
\usepackage{graphicx}
\usepackage{dcolumn}
\usepackage{natbib}
\usepackage{color}
\usepackage{xcolor}
\usepackage{multirow}
\usepackage{ulem}

\begin{document}


\title{Superluminal observers do not explain quantum superpositions}

\author{Andrzej Grudka}
\affiliation{Institute of Spintronics and Quantum Information, Faculty of Physics, Adam Mickiewicz University, 61-614 Pozna\'n, Poland}

\author{J\c{e}drzej Stempin}
\affiliation{Faculty of Physics, Adam Mickiewicz University, 61-614 Pozna\'n, Poland}

\author{Jan W{\'o}jcik}
\affiliation{Faculty of Physics, Adam Mickiewicz University, 61-614 Pozna\'n, Poland}

\author{Antoni W{\'o}jcik}
\affiliation{Institute of Spintronics and Quantum Information, Faculty of Physics, Adam Mickiewicz University, 61-614 Pozna\'n, Poland}

\date{\today}


\begin{abstract}
The quantum description of reality is quite different from the classical one. Understanding this difference at a fundamental level is still an interesting topic. Recently, Dragan and Ekert [New J. Phys. 22 (2020) 033038] postulated that considering so-called superluminal observers can be useful in this context. In particular, they claim that the full mathematical structure of the generalized Lorentz transformation may imply the emergence of multiple quantum mechanical trajectories. On the contrary, here we show that the generalized Lorentz transformation, when used in a consistent way, does not provide any correspondence between the classical concept of a definite path and the multiple paths of quantum mechanics.
\end{abstract}

\maketitle


{\it Introduction.} Quantum physics is full of counterintuitive phenomena. We can just accept this fact, admitting that our intuition is not based on laboratory experiments but on everyday (macro) experiences. Nevertheless  it is interesting to check if we can find for some, at first sight paradoxical, phenomena a more intuitive way of explanation. To be specific, let us consider the problem of the difference between classical and quantum physics with respect to the notion of particle trajectory. In classical physics, particles move along well-defined trajectories. On the other hand, quantum mechanics demands the abandonment of this notion and describes particles as moving along multiple paths. The double slit experiment is a paradigmatic example of such phenomena. Although particles can be in a superposition of two (or many) different positions when measured, they always appear at a particular position (do not split). Recently, Dragan and Ekert attempted to provide a perspective in which multiple paths of quantum mechanics appear naturally \cite{Dr}. They present a simple thought experiment with a single photon reflected from a mirror. After analyzing this experiment from the point of view of a superluminal observer, they conclude that multiple paths of quantum mechanics correspond to definite trajectories of classical mechanics. In their own words, "even if we start with an idea of a classical particle moving along a single path, it is only a matter of a change in the reference frame to arrive at a scenario involving more than one path."

\begin{figure}[t]
 \begin{tabular}{cc}
 \includegraphics[width=0.5\textwidth]{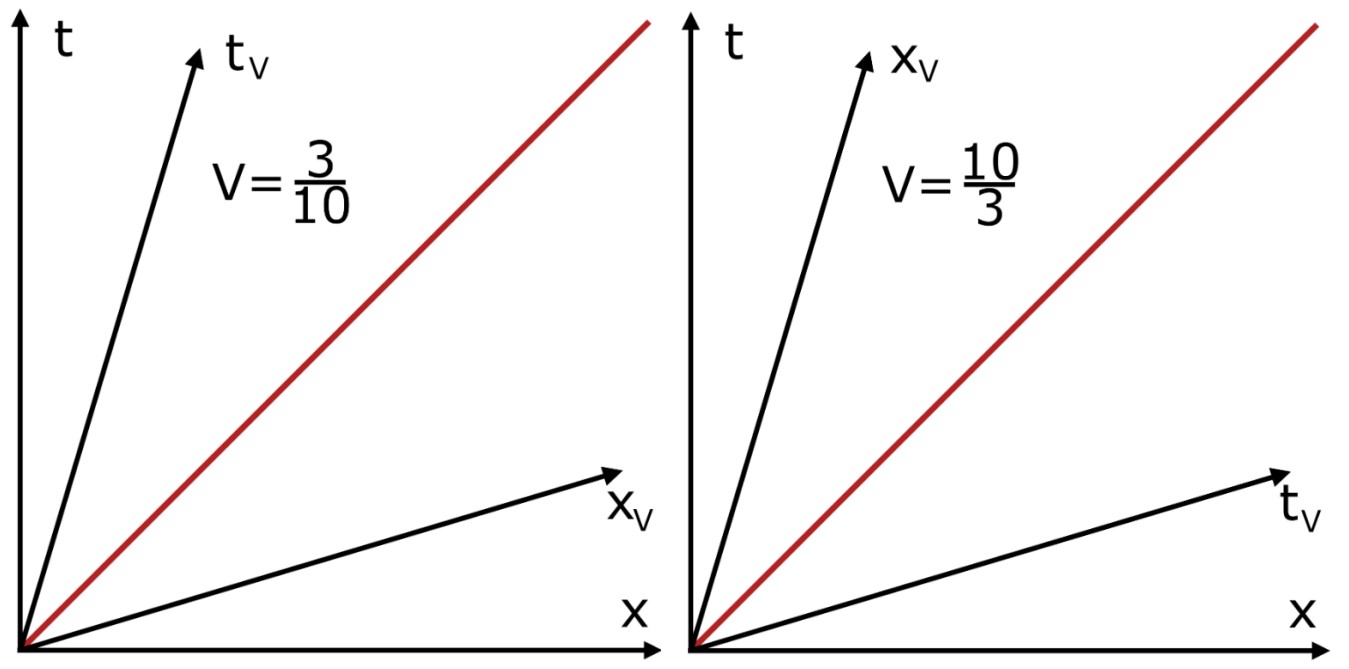}
 \end{tabular}
 \caption{Axes of coordinate systems and worldline of light ray. Coordinate system in rest $(t,x)$ and coordinate systems $(t_V,x_V)$ for $V=3/10$ (left) and $V=10/3$ (right).} 
\label{fig1}
\end{figure}

\begin{figure}[t]
 \begin{tabular}{cc}
 \includegraphics[width=0.5\textwidth]{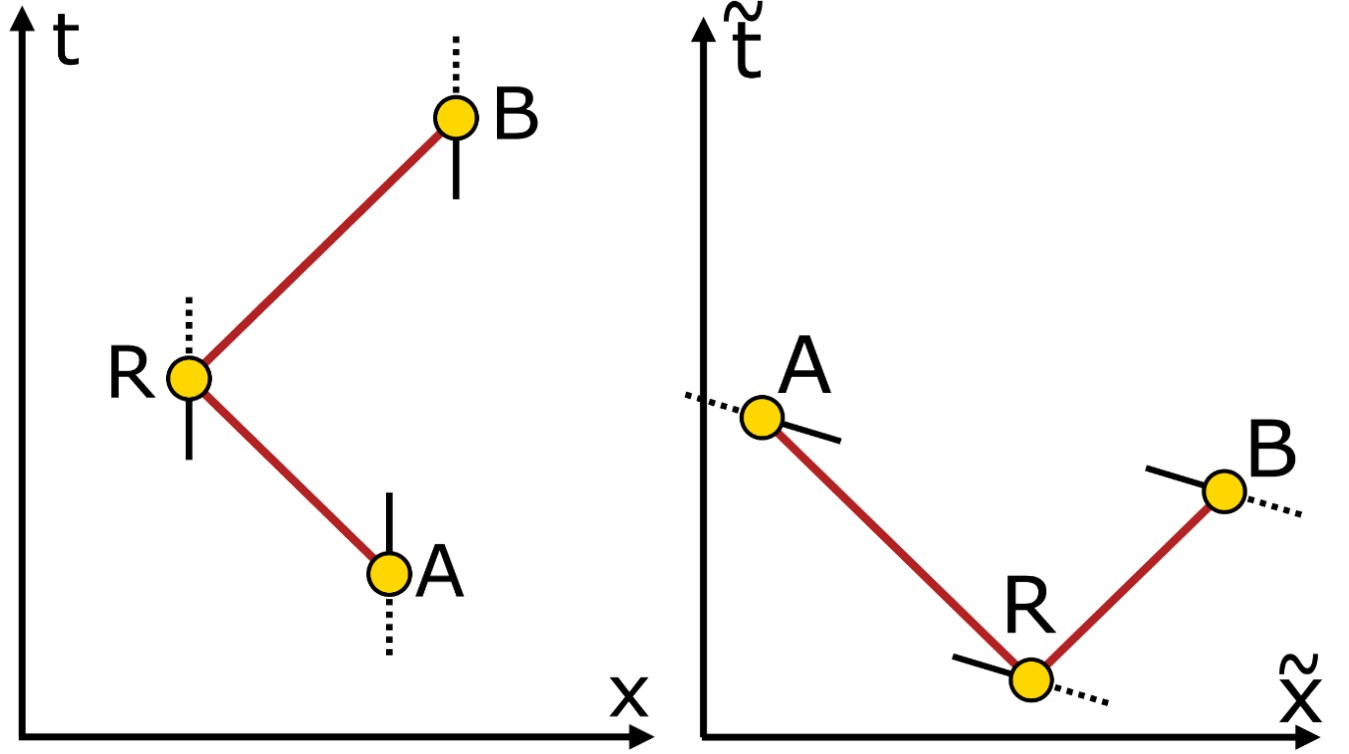}
 \end{tabular}
\caption{Left: process described by the observer $O$ as emission (event $A$), reflection (event $R$) and absorption (event $B$) of a photon. Right: the same process as seen by the observer $\tilde{O}$. A solid line denotes the world line of the TLA in $g$ state, whereas a dashed line denotes the world line of the TLA in $e$ state. Yellow (full) circles denote flipping of the TLA or mirror. We assume that the mass of the mirror and TLAs is large, so that in the scale of the picture, changes in their velocity after photon emission, reflection, or absorption are not visible.}
\label{fig2}
\end{figure}

In this note, we impair the above conclusion. We show that consistent analysis of the mentioned experiment in both subluminal and superluminal reference frames does not provide any explanation of the multiple paths expected by quantum mechanics. We note that interesting comments were also presented in \cite{hr} (see \cite{AA} for Dragan and Ekert reply).
\\\\
{\it Tools.}
To analyze the thought experiment of Dragan and Ekert in detail, we introduce the notion of two level atoms (TLAs), which can emit or absorb photons (for simplicity, we consider atoms and photons, but the argument is valid also for nonzero mass particles emitted by some appropriate systems). TLA has two internal states, which are denoted as $g$ (low energy state) and $e$ (high energy state). Both the absorption and the emission of photons are physical processes inevitably encoded on the TLA by the change of its internal state (as well as momentum change). Because what is absorption for one observer may be interpreted as emission for another, we will use the invariant term - state flip, which does not specify if the change was of the form $g \rightarrow e$ or $e \rightarrow g$. We will shortly say that TLA flips if it absorbs or emits a photon. For example, from the point of view of a subluminal observer, TLA goes from $g$ to $e$ and absorbs a photon, while for a superluminal observer, this event may appear in the following way: TLA goes from $g$ to $e$ and absorbs a photon (see event B in Fig. \ref{fig2}). We will also use another device, a mirror, which can reflect photons. We assume that mirror can be in one of two distinguishable states (e.g., momentum states) denoted similarly as $g$  and $e$, but on the contrary to the TLA case, these states are degenerated. While reflecting the photon, the mirror will flip (it changes its momentum without any significant energy cost), and one can check if the mirror reflected any photons by examining its state.
\\\\
{\it Coordinate systems and observers.}
As we consider two TLAs and a mirror that initially are at rest with respect to each other, it will be natural to use coordinates $(t,x)$ associated with the observer $O$ who is also at rest. We assume $c=1$ and measure both time and space in the same units.
We will also use coordinates $(t_V,x_V)$ defined as
\begin{equation}
    \begin{pmatrix}
        t_V\\x_V
    \end{pmatrix} = L_V\begin{pmatrix}
        t\\x
    \end{pmatrix},
\end{equation}
where 
\begin{equation}\label{eq}
    L_V = \pm \frac{1}{\sqrt{|1-V^2|}}\begin{pmatrix}
        1&&-V\\-V&&1
    \end{pmatrix}
\end{equation}
is valid transformation provided that $V \neq \pm 1$. In Eq. (\ref{eq}) $\pm$ we choose sign $+$ for $V<1$ and sign $-$ for $V>1$. With this choice, our transformation $L_V$ is the same as the transformations used by Dragan and Ekert (Eqs. 8 and 9 in their paper). Moreover, this choice guarantees that $L_V^{-1}=L_{-V}$. The above mentioned coordinate systems are presented in Fig. \ref{fig1}.  Note that for $|V|<1$ transformation, $L_V$ is just the usual Lorentz transformation. In the case of $|V|>1$ Dragan and Ekert called $L_V$ generalized Lorentz transformation. 
An invariant interval $ds^2=dt^2-dx^2$ is given by
\begin{equation}
    ds^2=dt_V^2-dx_V^2
\end{equation}
for $|V|<1$ and by
\begin{equation}
    ds^2=dx_V^2-dt_V^2
\end{equation}
for $|V|>1$.
Note that in the case of time separated events $ds^2 \geq 0$ their $t$-coordinate ordering is invariant with respect to $L_V$ with $|V|<1$ which is not guaranteed in the case of $|V|>1$. Anyway, it is obvious that coordinates are essentially irrelevant and that any process can be described in any coordinate system.

In the standard relativistic theory, there should be no particles that are at rest with respect to the coordinate system $(t_V,x_V)$ ($|V|>1$) which from now on will be denoted for simplicity as $(\tilde{t},\tilde{x})$ and so there should be no observers who consider $\tilde{t}$ and $\tilde{x}$ as time and space coordinates, respectively. Let us, however, try to generalize the standard approach and postulate the existence of an observer $\tilde{O}$ whose worldline is $\tilde{x}=\text{const.}$, i.e., in the $(t,x)$ coordinate system he moves with superluminal velocity and who uses $(\tilde{t},\tilde{x})$ as his coordinate system.
\\\\
{\it Analysis of the thought experiment.} 
Let us now analyze the process depicted in Fig. \ref{fig2} from both observers 
$O$ and $\tilde{O}$ perspectives. Below, we use capital letters to denote both spacetime events and their corresponding TLAs. From the point of view of an observer, $O$ the sequence of events is as follows. First, at time $t_A$ the TLA $A$ (initially prepared in state $e$) flips and emits a photon (event $A$). Then, at time $t_R$ mirror flips and reflects the photon (event $R$). Finally, the photon is absorbed (event $B$) at time $t_B$ by the TLA B (initially prepared in state $g$). 
This process is depicted in the left part of Fig. \ref{fig2} and corresponds to Fig. 3a in the Dragan and Ekert paper.

\begin{figure}[b]
 \begin{tabular}{cc}
 \includegraphics[width=0.5\textwidth]{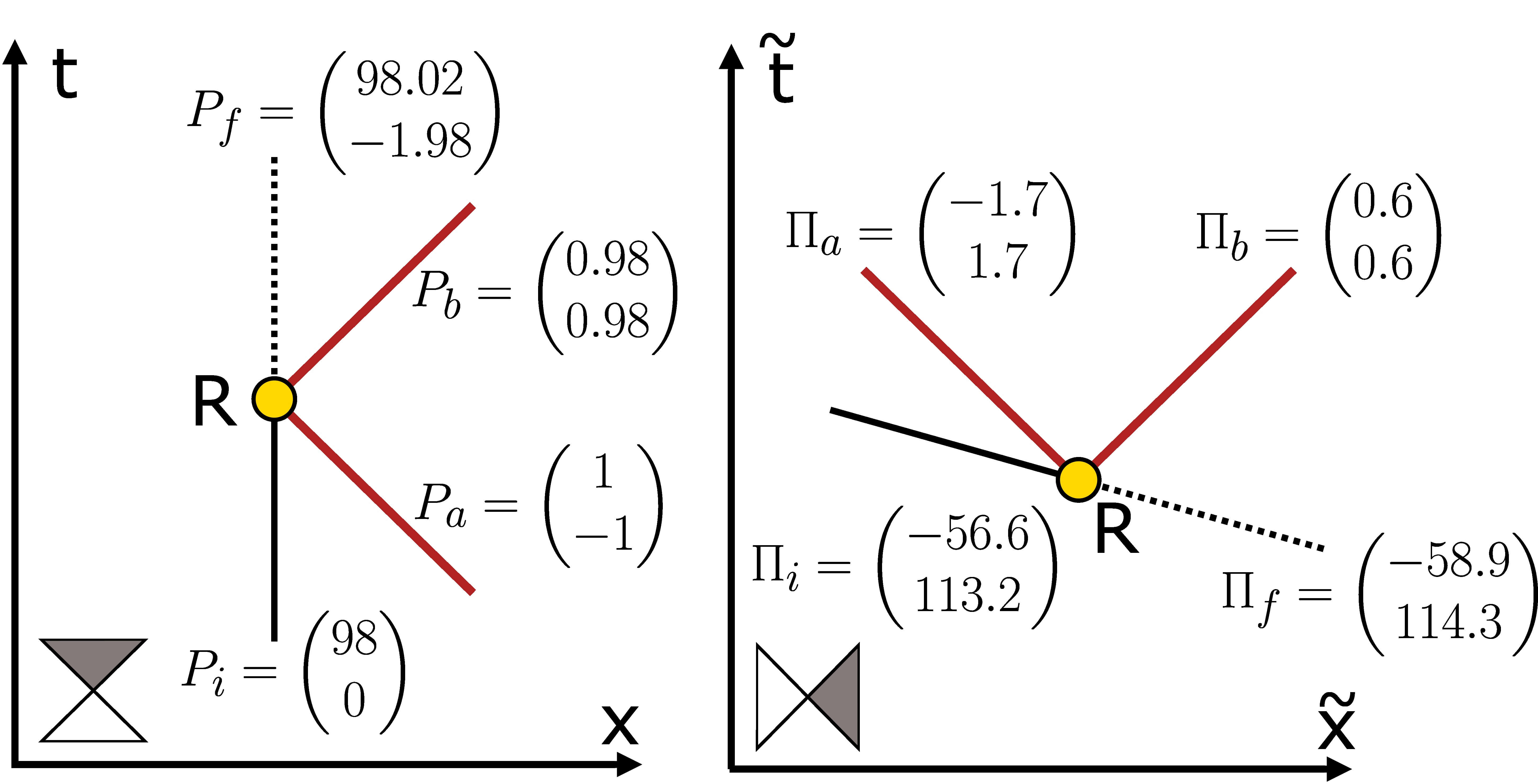}
 \end{tabular}
\caption{Reflection of the photon with initial energy $E_a=1$ by the mirror (initially at rest with mass $m=98$) as seen by observer $O$ (left part). The same process as seen by observer $\tilde O$ (right part). Components of energy-momentum vectors are presented. There are also included past (white) and future (gray) light cones.}
\label{ans}
\end{figure}

\begin{figure}[t]
 \begin{tabular}{cc}
 \includegraphics[width=0.5\textwidth]{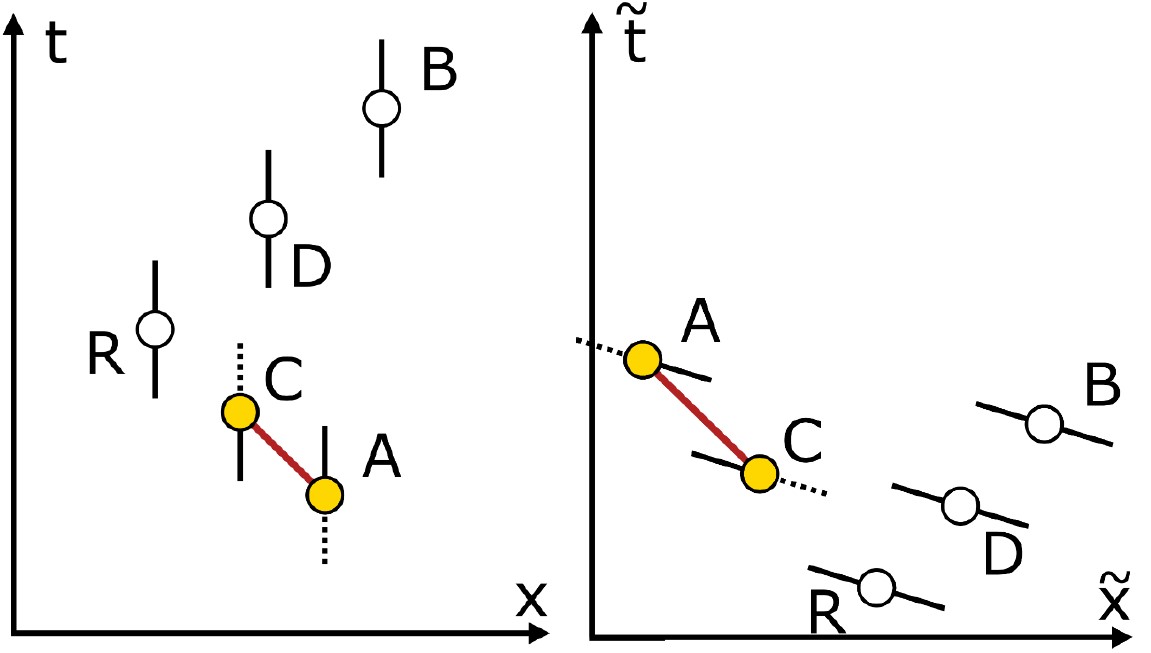}
 \end{tabular}
\caption{Left: process described by the observer $O$ as emission (event $A$), and absorption (event $C$) of a photon. Right: the same process as seen by the observer $\tilde{O}$. Yellow (full) circles denote flipping of the TLA or mirror, whereas white (empty) circles denote that there was no flipping.}
\label{fig3}
\end{figure}
\begin{figure}[b]
 \begin{tabular}{cc}
 \includegraphics[width=0.5\textwidth]{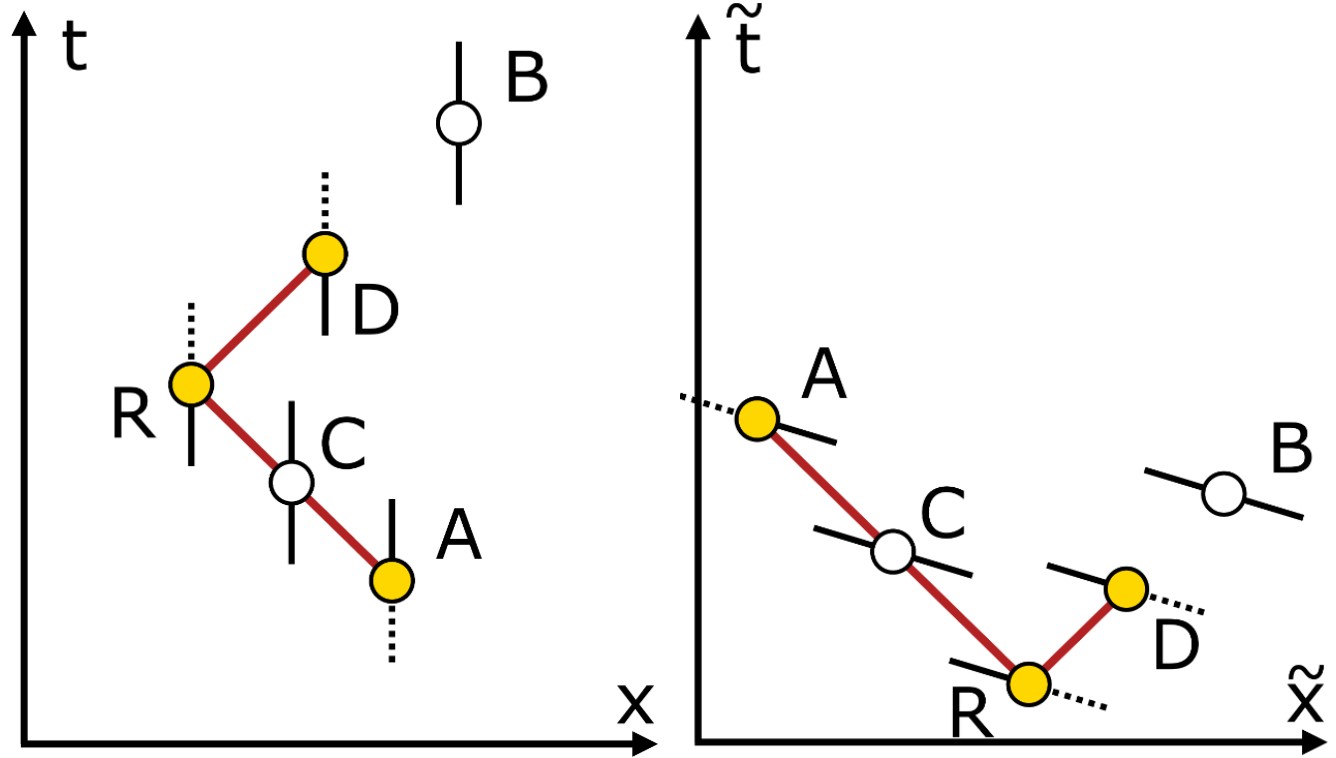}
 \end{tabular}
\caption{Left: process described by the observer $O$ as emission (event $A$), reflection (event $R$) and absorption (event $D$) of a photon. Right: the same process as seen by the observer $\tilde{O}$.}
\label{fig4}
\end{figure}
How this process can be described by an observer $\tilde{O}$? The sequence of events (ordered with the use of $\tilde{t}$ coordinate) looks like this. First at time $\tilde{t}_R$ mirror flips. Then, at time $\tilde{t}_B$ the TLA $B$ flips, and finally, at time $\tilde{t}_A$  TLA $A$ flips (see the right part of Fig. \ref{fig2}). This sequence of events can be interpreted by the observer $\tilde{O}$ with respect to the flow of the $\tilde{t}$ coordinate in the following way. When the mirror flips, it emits a pair of photons, which then cause the flips of both TLAs. It is clear that our analysis of the mentioned process from the point of view of the observer $\tilde{O}$ is in contradiction with the interpretation of Dragan and Ekert. They propose that at the event $R$, mirror emits not a pair of photons but a single one. This single photon is claimed to be in superposition of the two paths. However, such a single photon is not able to induce observable effects in both paths (as long as the usual quantum mechanical notion of superposition is involved), so it cannot explain the occurrence of events $A$ and $B$.

{\it Detailed description of reflection in two different reference frames.} One can raise some objections about consistency between both points of view on the event $R$. This is because reflection observed by $O$ can happen without changing the internal energy of the mirror, whereas the emission of two photons observed by $\tilde{O}$ seems to require that the internal energy of the mirror changes. To resolve this apparent paradox one can analyze energy-momentum two-vectors describing systems involved in that process. Let energy-momentum vectors be denoted as $P_i$ ($P_f$) for the initial (final) state of the mirror and $P_a$ ($P_b$) for the initial (final) state of the photon. We find it instructive to consider here the example with specific numerical values. Such an example is presented in Fig. \ref{ans}. The left part of this picture represents the subluminal observer's $O$ point of view, with the mirror of rest mass $m=98$ initially at rest and the photon initially moving to the left with energy $E_a=1$. Thus energy-momentum vectors are initially given by $P_i=\begin{pmatrix} 98\\0\end{pmatrix}$ and $P_a=\begin{pmatrix} 1\\-1\end{pmatrix}$. Energy-momentum conservation
\begin{equation}\label{ep1}
    \begin{pmatrix}
        98\\0
    \end{pmatrix}+
    \begin{pmatrix}
       1\\-1
    \end{pmatrix}=
    \begin{pmatrix}
       98.02\\-1.98
    \end{pmatrix}+
    \begin{pmatrix}
      0.98\\0.98
    \end{pmatrix}
\end{equation}
leads to the energy-momentum vectors $P_f=\begin{pmatrix} 98.02\\-1.98\end{pmatrix}$ and $P_b=\begin{pmatrix} 0.98\\0.98\end{pmatrix}$  after the event $R$. One can check that the change in internal energy (rest mass) of the mirror $
    \Delta E_{int}=\sqrt{P_f^2}-\sqrt{P_i^2}$ is indeed zero.

The right part of Fig \ref{ans} represents the same process in coordinates used by a superluminal observer who moves with the speed $V=2$. Transformation to superluminal coordinates (Eq. \ref{eq})
\begin{equation}
    L_2 = \frac{1}{\sqrt{3}}\begin{pmatrix}
        -1&&2\\2&&-1
    \end{pmatrix}.
\end{equation}
applied to Eq. \ref{ep1} gives
\begin{equation}\label{ep2}
    \begin{pmatrix}
        -56.6\\113.2
    \end{pmatrix}+
    \begin{pmatrix}
       -1.7\\1.7
    \end{pmatrix}=
    \begin{pmatrix}
       -58.9\\114.3
    \end{pmatrix}+
    \begin{pmatrix}
      0.6\\0.6
    \end{pmatrix}.
\end{equation}
Energy-momentum vectors after the transformation are denoted by $\Pi_j= L_2 P_j$
(with $j \in \{i, f, a, b\}$). Rounded values of the components of these energy-momentum vectors are given in right part of Fig. \ref{ans}. One can check (note that transformation $L_2$ changes the sign of metric tensor) that the change in the internal energy of the device as seen by the superluminal observer is again zero. Note, that this must be so because the square of the energy-momentum vector is an invariant scalar.
Thus, there is no inconsistency. There is no change in internal energy during reflection for any (subluminal or superluminal) observer. However, the superluminal observer sees an event in which worldlines of two photons are created. Hence, this process looks like the emission of two photons, but without a change of the internal energy of the mirror.
Let us also emphasize another point.
It is clear that it is impossible to derive energy-momentum conservation given by Eq. \ref{ep2} by comparing the sum of energy-momentum vectors before event $R$ and after it, if "before" and "after" are defined according to the transformed time coordinate $\tilde t$. To make it clear, we put in Fig. \ref{ans} past and future light cones. Now one can see that the proper statement of the energy-momentum conservation in both coordinate systems is to sum energy-momentum vectors which are in the past of the $R$ and compare it to the sum of energy-momentum vectors which are in its future. Note also, that by looking at the orientation of the past-future cones in the right part of Fig. \ref{ans} one can recognize that, although the process appears as an emission of two photons, it is indeed the process of reflection.

{\it Additional detectors.} Dragan and Ekert also analyzed the behavior of two detectors - one in the path between $A$ and $R$ and another in the path between $B$ and $R$. So we use two additional TLAs as these detectors (TLA $C$ and TLA $D$ in Figs. \ref{fig3} and \ref{fig4}). From a point of view of the observer $O$ it is obvious that both TLAs $C$ and $D$ cannot both flip. Dragan and Ekert argue that this observation, translated into the superluminal observer's point of view, can explain why a quantum particle cannot be found in two positions at once despite being in appropriate superposition. So let us present the process by which a photon is detected by the TLA $C$ in Fig. \ref{fig3} and by the TLA $D$ in Fig. \ref{fig4}. Note that both TLAs serving as detectors are initially (from a point of view of the observer $O$) in the low energy state $g$. The observer $\tilde{O}$ sees (right part of Fig. \ref{fig3}) process in which the initially excited TLA $C$ flips and emits a photon, which is then absorbed by TLA $A$ which also flips. In this case, mirror, TLA $B$ and TLA $D$ don't flip. The observer $\tilde{O}$ interpretation of the process depicted in Fig. \ref{fig4} is following. First, at the time $(\tilde{t}_R)$ mirror flips and emits a pair of photons. One of them is absorbed by the TLA $D$ which flips, and the second one is absorbed by the TLA $A$ which also flips. TLA $C$ doesn't flip not because a photon was registered at TLA $D$ but because the other photon was transmitted to TLA $A$.

{\it Conclusions.} 
It was shown that the reflection of a single photon (following a definite trajectory) from the mirror does not correspond in the superluminal coordinate system to the superposition of a single photon in two paths. A superluminal observer sometimes sees two photons instead of one. However, each of these photons follows a definite trajectory.

\bibliographystyle{plain}

\end{document}